\documentclass[11pt]{article}
\usepackage{epsfig}
\usepackage{a4}
\usepackage{amssymb}
\usepackage{lineno}
 \usepackage[frenchb,english]{babel}
 \usepackage{textcomp}
 \usepackage{amssymb}
 \usepackage{currvita}
 \usepackage{multirow}
 \usepackage[applemac]{inputenc}
\usepackage{fancyvrb}
\usepackage{graphics}
\usepackage{subfigure}

\linenumbers
\newcommand{\ANKARA}      {29}
\newcommand{\ANNECY}      {10}
\newcommand{\BARI}        {34}
\newcommand{\BARIINFN}    {20}
\newcommand{\BERN}        {4}
\newcommand{\BOLOGNA}     {14}
\newcommand{\BOLOGNAINFN} {15}
\newcommand{\BRUSSELS}    {38}

\newcommand{\DUBNA}      {22}
\newcommand{\FRASCATI}   {8}
\newcommand{\FUNABASHI}  {26}	
\newcommand{\HAIFA}      {28}
\newcommand{\HAMBURG}    {11}
\newcommand{\GAZWADONG}  {35}
\newcommand{\KARIYA}     {32}
\newcommand{\KOBE}       {3}
\newcommand{\LAQUILA}    {21}
\newcommand{\LNGS}       {18}
\newcommand{\LYON}       {5}
\newcommand{\MOSCOWINR}  {1}
\newcommand{\MOSCOWITEP} {23}
\newcommand{\MOSCOWLPI}  {7}
\newcommand{\MOSCOWSINP} {2}
\newcommand{\MUNSTER}    {24}
\newcommand{\NAGOYA}     {25}
\newcommand{\NAPOLI}     {19}
\newcommand{\NAPOLIINFN} {16}
\newcommand{\OBNINSK}    {27}
\newcommand{\PADOVA}     {13}
\newcommand{\PADOVAINFN} {9}
\newcommand{\ROMA}       {36}
\newcommand{\ROSTOCK}    {30}
\newcommand{\SALERNO}    {12}
\newcommand{\SOFIA}      {33}
\newcommand{\STRASBOURG} {17}
\newcommand{\UTSUNOMIYA} {37}
\newcommand{\ZAGREB}     {31}
\newcommand{\ZURICH}     {6}
\newcommand{\CORR}       {*}

\newcommand{\OperaInstitutes}{
\MOSCOWINR   . INR-Institute for Nuclear Research of the Russian Academy of Sciences, RUS-117312 Moscow, Russia\\
\MOSCOWSINP  . SINP MSU-Skobeltsyn Institute of Nuclear Physics of Moscow State University, RUS-119992 Moscow, Russia \\
\KOBE        . Kobe University, J-657-8501 Kobe, Japan \\
\BERN        . Centre for Research and Education in Fundamental Physics, Laboratory for High Energy Physics (LHEP),
               University of Bern, CH-3012 Bern, Switzerland \\
\LYON        . IPNL, Universit\'e Claude Bernard Lyon 1, CNRS/IN2P3, F-69622 Villeurbanne, France\\
\ZURICH      . ETH Zurich, Institute for Particle Physics, CH-8093 Zurich, Switzerland \\
\MOSCOWLPI   . LPI-Lebedev Physical Institute of the Russian Academy of Sciences, RUS-117924 Moscow, Russia\\
\FRASCATI    . INFN - Laboratori Nazionali di Frascati dell'INFN, I-00044 Frascati (Roma), Italy \\
\PADOVAINFN  . INFN Sezione di Padova, I-35131 Padova, Italy \\
\ANNECY      . LAPP, Universit\'e de Savoie, CNRS/IN2P3, F-74941 Annecy-le-Vieux, France\\
\HAMBURG     . Hamburg University, D-22761 Hamburg, Germany\\
\SALERNO     . Dipartimento di Fisica dell'Universit\`a  di Salerno and INFN, I-84084 Fisciano, Salerno, Italy \\
\PADOVA      . Dipartimento di Fisica dell'Universit\`a  di Padova, I-35131 Padova, Italy \\
\BOLOGNA     . Dipartimento di Fisica dell'Universit\`a  di Bologna, I-40127 Bologna, Italy \\
\BOLOGNAINFN . INFN Sezione di Bologna, I-40127 Bologna, Italy \\
\NAPOLIINFN  . INFN Sezione di Napoli, 80125 Napoli, Italy \\
\STRASBOURG  . IPHC, Universit\'e de Strasbourg, CNRS/IN2P3, F-67037 Strasbourg, France\\
\LNGS        . INFN - Laboratori Nazionali del Gran Sasso, I-67010 Assergi (L'Aquila), Italy \\
\NAPOLI      . Dipartimento di Fisica dell'Universit\`a Federico II di Napoli, 80125 Napoli, Italy \\
\BARIINFN    . INFN Sezione di Bari, I-70126 Bari, Italy \\
\LAQUILA     . Dipartimento di Fisica dell'Universit\`a dell'Aquila and INFN, I-67100 L'Aquila,
Italy \\
\DUBNA       . JINR-Joint Institute for Nuclear Research, RUS-141980 Dubna, Russia \\
\MOSCOWITEP  . ITEP-Institute for Theoretical and Experimental Physics, RUS-117259 Moscow, Russia \\
\MUNSTER     . University of M\"unster, D-48149 M\"unster, Germany\\
\NAGOYA      . Nagoya University, J-464-8602 Nagoya, Japan\\
\FUNABASHI   . Toho University, J-274-8510 Funabashi, Japan \\
\OBNINSK     . Obninsk State University, Institute of Nuclear Power Engineering, RUS-249020 Obninsk, Russia\\
\HAIFA       . Department of Physics, Technion, IL-32000 Haifa, Israel\\
\ANKARA      . METU-Middle East Technical University, TR-06531 Ankara, Turkey \\
\ROSTOCK     . Fachbereich Physik der Universit\"at Rostock, D-18051 Rostock, Germany\\
\ZAGREB      . IRB-Rudjer Boskovic Institute, HR-10002 Zagreb, Croatia\\
\KARIYA      . Aichi University of Education, J-448-8542 Kariya (Aichi-Ken), Japan\\
\SOFIA       . Faculty of Physics, Sofia University ``St. Kliment Ohridski", BG-1000 Sofia, Bulgaria\\
\BARI        . Dipartimento di Fisica dell'Universit\`a  di Bari, I-70126 Bari, Italy \\
\GAZWADONG   . Gyeongsang National University, 900 Gazwa-dong, Jinju 660-300, Korea\\
\ROMA        . Dipartimento di Fisica dell'Universit\`a  di Roma ``La Sapienza" and INFN, I-00185 Roma, Italy \\
\UTSUNOMIYA  . Utsunomiya University, J-321-8505 Tochigi-Ken, Utsunomiya, Japan\\
\BRUSSELS    . IIHE, Universit\'e Libre de Bruxelles, B-1050 Brussels, Belgium \\
\CORR        . Corresponding Author\\
}

\newcommand{\OperaAuthorList}{
N.~Agafonova$^{\MOSCOWINR}$,
A.~Anokhina$^{\MOSCOWSINP}$,
S.~Aoki$^{\KOBE}$,
A.~Ariga$^{\BERN}$,
T.~Ariga$^{\BERN}$,
L.~Arrabito$^{\LYON}$,
D.~Autiero$^{\LYON}$,
A.~Badertscher$^{\ZURICH}$,
A.~Bagulya$^{\MOSCOWLPI}$,
F.~Bersani~Greggio$^{\FRASCATI}$
% ...
\\
}

\begin{document}

\def\numunue{\nu_\mu\rightarrow\nu_e}
\def\numunutau{\nu_\mu\rightarrow\nu_\tau}
\def\nuebar{\bar\nu_e}
\def\nue{\nu_e}
\def\nutau{\nu_\tau}
\def\numubar{\bar\nu_\mu}
\def\numu{\nu_\mu}
\def\ra{\rightarrow}
\def\numubarnuebar{\bar\nu_\mu\rightarrow\bar\nu_e}
\def\nuebarnumubar{\bar\nu_e\rightarrow\bar\nu_\mu}
\def\osc{\rightsquigarrow}

\def\inteni{{\cal I}_{pot}}
\def\fmerit{{\cal F}}

\title{\bf The detection of neutrino interactions in the emulsion/lead target of the OPERA experiment}

\maketitle

\Large
\begin{center}
Submitted to JINST

\end{center}

\small
\author{\noindent \\ \OperaAuthorList }

\begin{flushleft}
\footnotesize{\OperaInstitutes }
\end{flushleft}



\vspace{0.2cm}

\begin{abstract}
The OPERA neutrino detector in the underground Gran Sasso Laboratory (LNGS) was designed to perform the first detection of neutrino
oscillations in appearance mode through the study of $\nu_\mu\rightarrow\nu_\tau$ oscillations. The apparatus consists of an emulsion/lead target complemented by electronic detectors and it is placed in the high energy long-baseline CERN to LNGS beam
(CNGS) 730 km away from the neutrino source. Runs with CNGS neutrinos were successfully carried out in 2007 and 2008 with the detector fully operational with its related facilities for the emulsion handling and analysis. After a brief description of the beam and of the experimental setup we report on the collection, reconstruction and analysis procedures of first samples of neutrino interaction events.

\end{abstract}

\section{Introduction}
Neutrino oscillations were anticipated nearly 50 years ago~\cite{osc} but they have been unambiguously observed only recently. Several experiments carried out in the last decades with atmospheric and accelerator neutrinos, as well as with solar and reactor neutrinos, contributed to our present understanding of neutrino mixing (see e.g.~\cite{review} for a review).


As far as the atmospheric neutrino sector is concerned, accelerator experiments can probe the same oscillation parameter region as atmospheric neutrino experiments~\cite{allexp}. This is the case of the OPERA experiment that has the main scientific task of the first direct detection of $\nu_\mu\rightarrow\nu_\tau$ appearance, an important missing tile in the oscillation scenario~\cite{Ereditato:1997qe,Acquafredda:2006ki,detpap}.


OPERA uses the long-baseline (L=730 km) CNGS neutrino beam~\cite{cngs} from CERN to LNGS, the largest underground physics laboratory in the world.
The challenge of the experiment is to measure the appearance of $\nutau$ from $\numu$ oscillations in an almost pure muon-neutrino beam.
Therefore, the detection of the short-lived $\tau$ lepton (c$\tau$ = 87.11 $\mu$m) produced in the charged-current (CC) interaction of a $\nu_\tau$ is mandatory. This sets two conflicting requirements: a large target mass to collect enough statistics and an extremely high spatial accuracy to observe the short-lived $\tau$ lepton.

The $\tau$ is identified by the detection of its characteristic decay topologies either in one prong (electron, muon or hadron) or in three-prongs;
 its short track is measured with a large mass target made of 1 mm thick lead plates (target mass and absorber material) interspaced with thin nuclear emulsion films (high-accuracy tracking devices). This detector is historically called Emulsion Cloud Chamber (ECC). Among past applications it was successfully used in the DONUT experiment for the first direct observation of the $\nutau$~\cite{donut}.

\begin{figure}
\begin{center}
  \includegraphics[width=15cm]{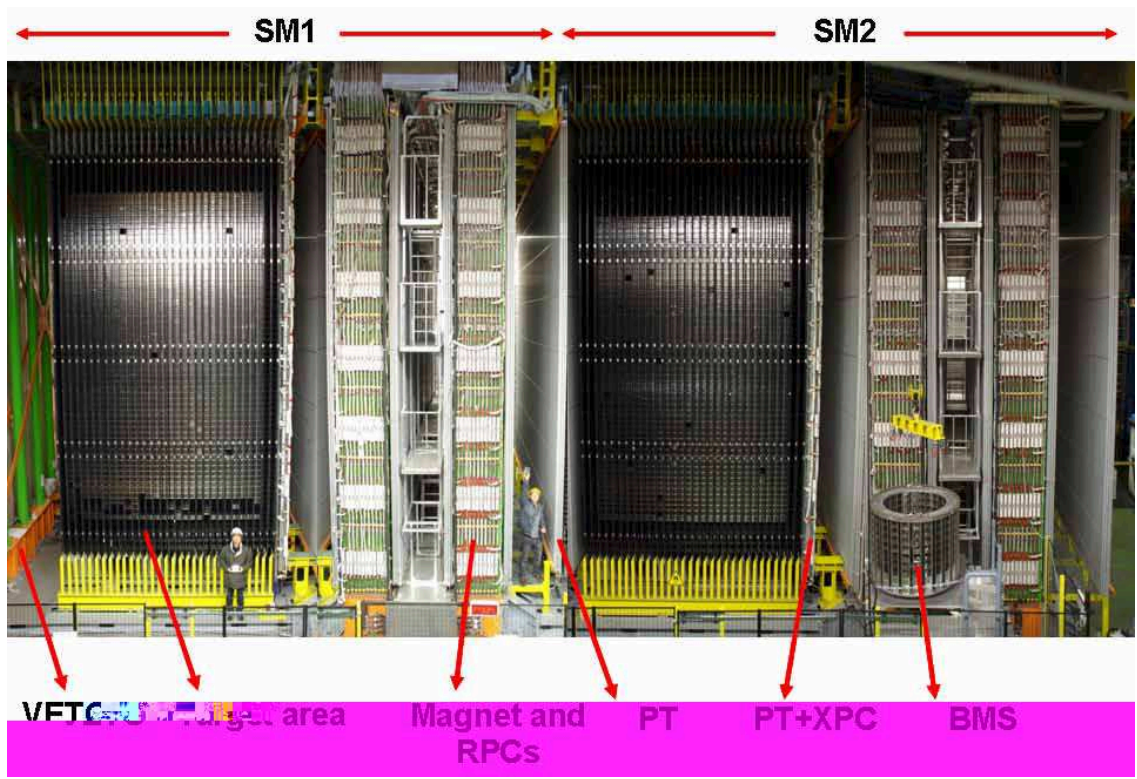}
   \caption{View of the OPERA detector. The upper horizontal lines indicate the position of the two identical supermodules (SM1 and SM2). The "target area" is made of walls filled with ECC bricks interleaved with planes of plastic scintillators (TT). Arrows show the position of the VETO planes, the drift tubes (PT) pulled alongside  the XPC, the
 magnets and the RPC installed between the magnet iron slabs. The Brick Manipulator System (BMS) is also visible. See \cite{detpap} for more details.} \label{detector}
\end{center}
\end{figure}

OPERA is a hybrid detector made of two identical Super Modules (SM) each consisting of a target section of about 625 tons made of emulsion/lead ECC modules (hereafter called ''bricks''), of a scintillator tracker detector (TT) needed to trigger the read-out and localize neutrino interactions within the target, and of a muon spectrometer (Figure \ref{detector}). The detector is equipped with an automatic machine (the Brick Manipulator System, BMS) that allows the online removal of bricks from the detector. Ancillary, large facilities are used for the handling, the development and the scanning of the emulsion films. Emulsion scanning is performed with two different types of automatic microscopes: the European Scanning System (ESS) \cite{ees,ees2} and the Japanese S-UTS \cite{suts}.

A target brick consists of 56 lead plates of 1 mm thickness interleaved with 57 emulsion films \cite{emulsions}. The plate material is a lead alloy with a small calcium content to improve its mechanical properties~\cite{lead}. The transverse dimensions of a brick are 12.8 $\times$ 10.2 cm$^2$ and the thickness along the beam direction is 7.9 cm (about 10 radiation lengths). The bricks are housed in a light support structure placed between consecutive TT walls. More details on the detector and on the ancillary facilities are given in \cite{detpap}.

\begin{figure}
\begin{center}
  \includegraphics[width=7cm]{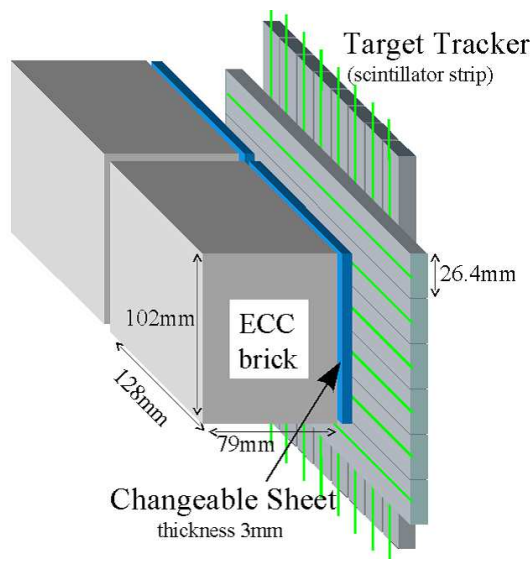}
   \caption{Schematic view of two bricks with their Changeable Sheets and target tracker planes.} \label{CSTTlayout}
\end{center}
\end{figure}

In order  to reduce the emulsion scanning load the use of Changeable Sheets (CS) film interfaces~\cite{CSD}, successfully applied in the CHORUS experiment~\cite{chorusCS}, was extended to OPERA. Tightly packed doublets of emulsion films are glued to the downstream face of each brick and can be removed without opening the brick. The global layout of brick, CS and TT is schematically shown in Figure \ref{CSTTlayout}.

Charged particles from a neutrino interaction in the brick cross the CS and produce a signal in the TT scintillators.
The corresponding brick is then extracted and the CS developed and analyzed in the scanning facilities at LNGS and in Nagoya.
The information of the CS is then used for a precise prediction of the position of the tracks
in the most downstream films of the brick, hence guiding the {\em scan-back} vertex finding procedure.

\begin{figure}
\begin{center}
  \includegraphics[width=15cm]{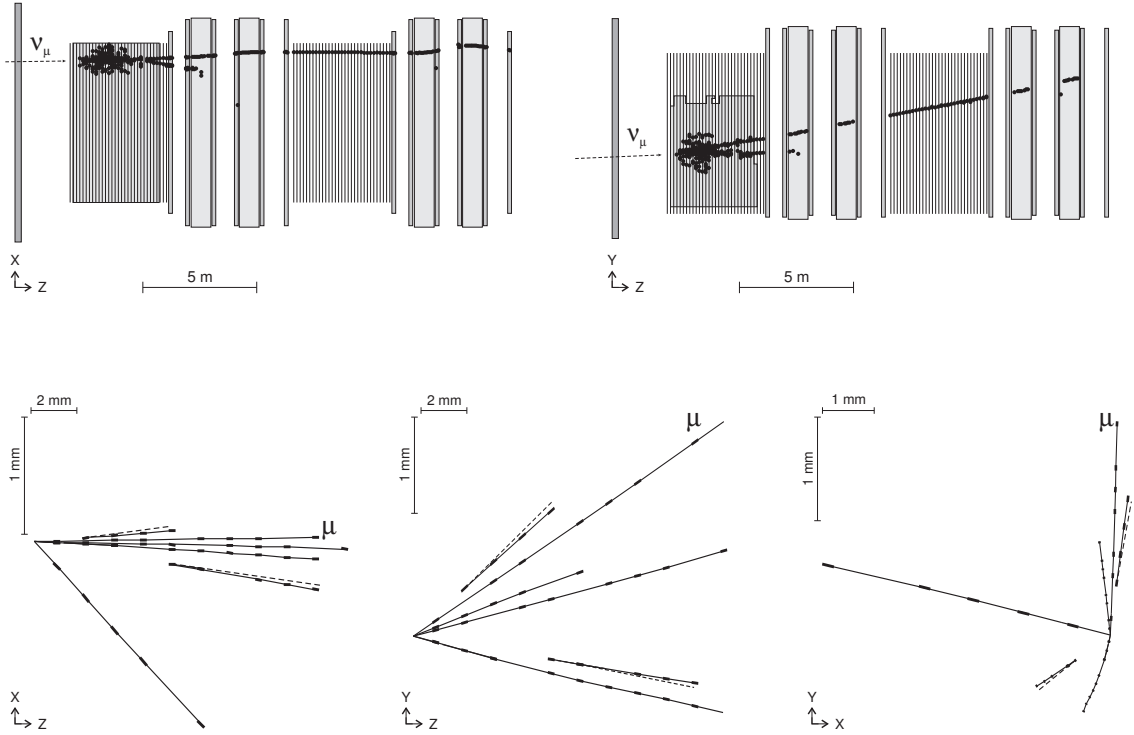}
   \caption{Top panels: on line display of an event seen by the OPERA electronic detectors (side and top views): a $\nu_\mu$ interacts in one of the first bricks of the first supermodule (SM) yielding hadrons and a muon which is detected in both SMs and whose momentum is measured by the magnets of the two SMs. Bottom panels: the vertex of the same event observed in the emulsion films (side, top and front views). Note the two $\gamma\rightarrow e^{+}e^{-}$ vertices: the opening angle between them is about 300 mrad. By measuring
  the energy of the $\gamma$'s one obtains a reconstructed invariant mass of $110\pm30$ MeV/$c^2$, consistent with the $\pi^{0}$ mass.} \label{babyopera}
\end{center}
\end{figure}

A reconstructed CC event is shown in the bottom panels of Figure \ref{babyopera}. In this case the detached event dimensions are of the order of a few millimeters, to be compared with the $\sim10$ m scale of the whole event reconstructed with the electronic detectors (top panels of Figure \ref{babyopera}).

First neutrino data were collected by OPERA in 2006 \cite{Acquafredda:2006ki} with the electronic detectors alone, and then in 2007 and 2008, for the first time with target bricks installed.  All steps from the prediction of the brick where the interaction occurred down to the kinematical analysis of the neutrino interactions are described in the following using as benchmark a sub-sample of the statistics accumulated during the CNGS runs.
The procedure has proven to be successful. We are presently in the process of a quantitative evaluation of the different experimental efficiencies that are involved in the analysis procedure, profiting of the increasing statistics of the reconstructed neutrino events.

\section {Real time detection of the CNGS Beam}

The CNGS neutrino beam \cite{cngs} was designed and optimized for the study of $\numunutau$ oscillations in appearance mode
by maximizing the number of CC $\nutau$  interactions at the LNGS site. After a short commissioning run in 2006 the CNGS operation started on September 2007 at rather low intensity. The first event inside the OPERA target was observed on October 3$^{rd}$.  Unfortunately, due to a fault of the CNGS facility, the physics run lasted only a few days. During this run $0.082\times10^{19}$ protons on target (p.o.t.) were accumulated with a mean value of $1.8\times10^{13}$  protons per extraction\footnote[1]{The 400 GeV proton beam is extracted from the CERN SPS in two 10.5 $\mu$s pulses, with design intensity of $2.4\times10^{13}$ p.o.t.}: this corresponds to about $\sim3.6$ effective nominal days of running. With such an integrated intensity 32 neutrino interactions in the bricks and 3 in the scintillator material of the target tracker were expected; we actually observed 38 events on time with the arrival of the beam at Gran Sasso.

\begin{figure}
\begin{center}
  \includegraphics[width=9cm]{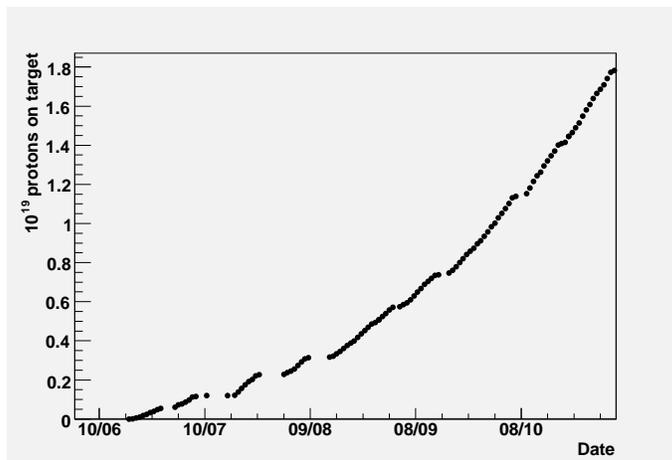}
   \caption{Integrated number of protons on target (p.o.t.) as a function of time for the 2008 CNGS run (June-November).} \label{pot2008}
\end{center}
\end{figure}

A much longer run took place in 2008 when $1.782\times10^{19}$ protons were delivered on the CNGS target. OPERA collected 10100 events on time  and among them 1700 interactions in the bricks. The other events originated outside the target region (spectrometers, supporting structures, rock surrounding the cavern, hall structures, etc.). The run featured a poor initial efficiency of the CERN complex, with an average of about 40\% until reaching an average  value of about 60\%. From October to November OPERA gathered the same number of events as from June to August. The 2008 CNGS integrated p.o.t. intensity as a function of time is shown in Figure~\ref{pot2008}.

During the 2007 and 2008 runs all electronic detectors were operational and the live time of the data acquisition system exceeded 99\%. More than 10 million events were collected by applying a minimum bias filter. The selection of beam related events relies upon a time stamp, based on the time synchronization accuracy of 100 ns between the CERN
beam GPS tagging and the OPERA timing system.

An automatic classification algorithm provides high efficiency in the selection of neutrino interactions inside the OPERA target both for CC and neutral-current (NC) events at the expenses of a slight contamination of neutrino interactions in the external material.

For the early 2007 run the algorithm selected 53 events occurring inside the target while expecting 50; the contamination from neutrino interactions outside the target amounted to $37\%$ (Monte Carlo simulation). The low purity of the selected event sample was due to some sub-detectors still being in the commissioning phase and to the OPERA target that was only partially filled with bricks. In the 2008 run $1663$ events were classified as interactions in the target (expected $1723$) with a contamination from outside events of only $7\%$.

\begin{figure}
\begin{center}
  \includegraphics[width=7.5cm]{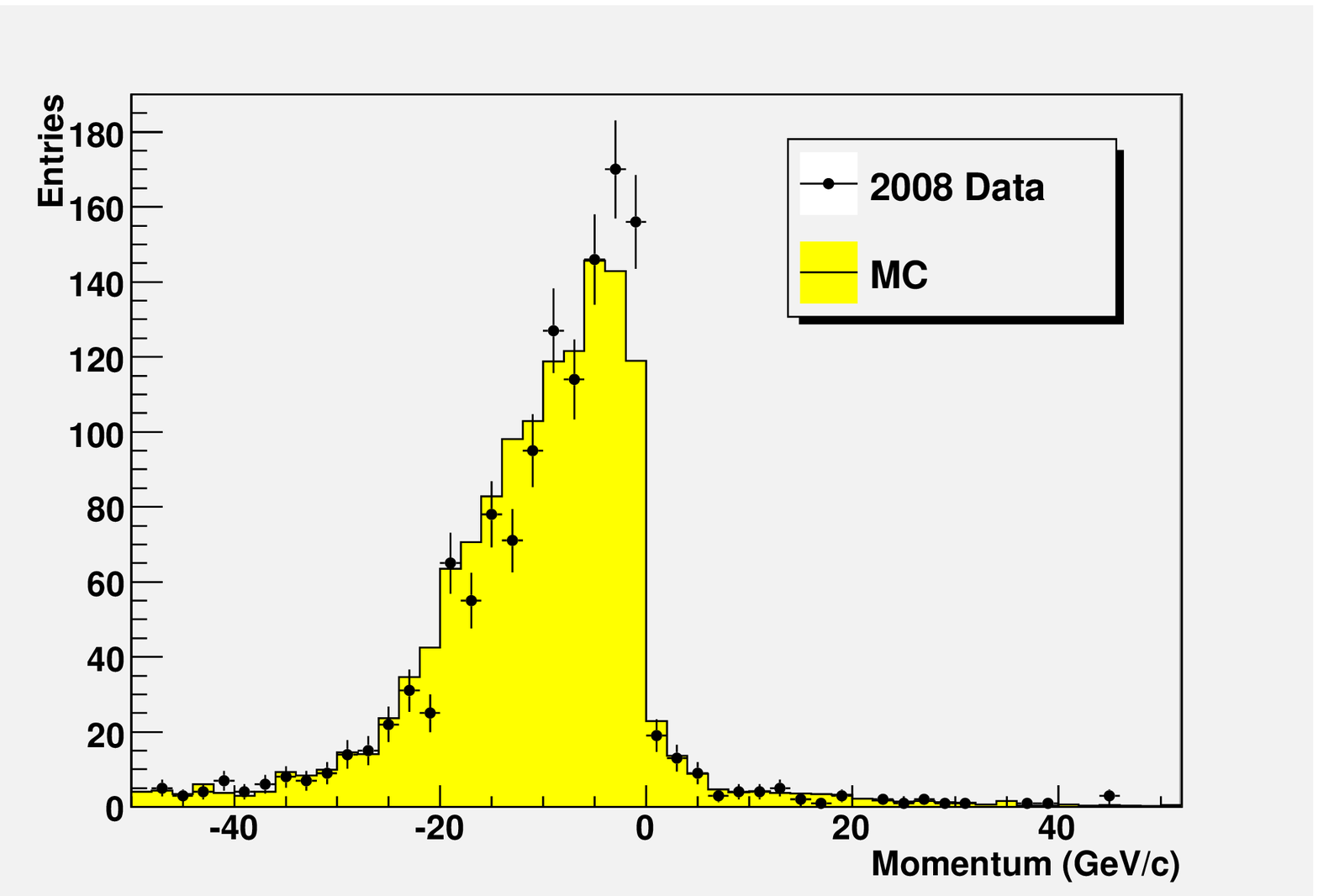}
  \includegraphics[width=7.5cm]{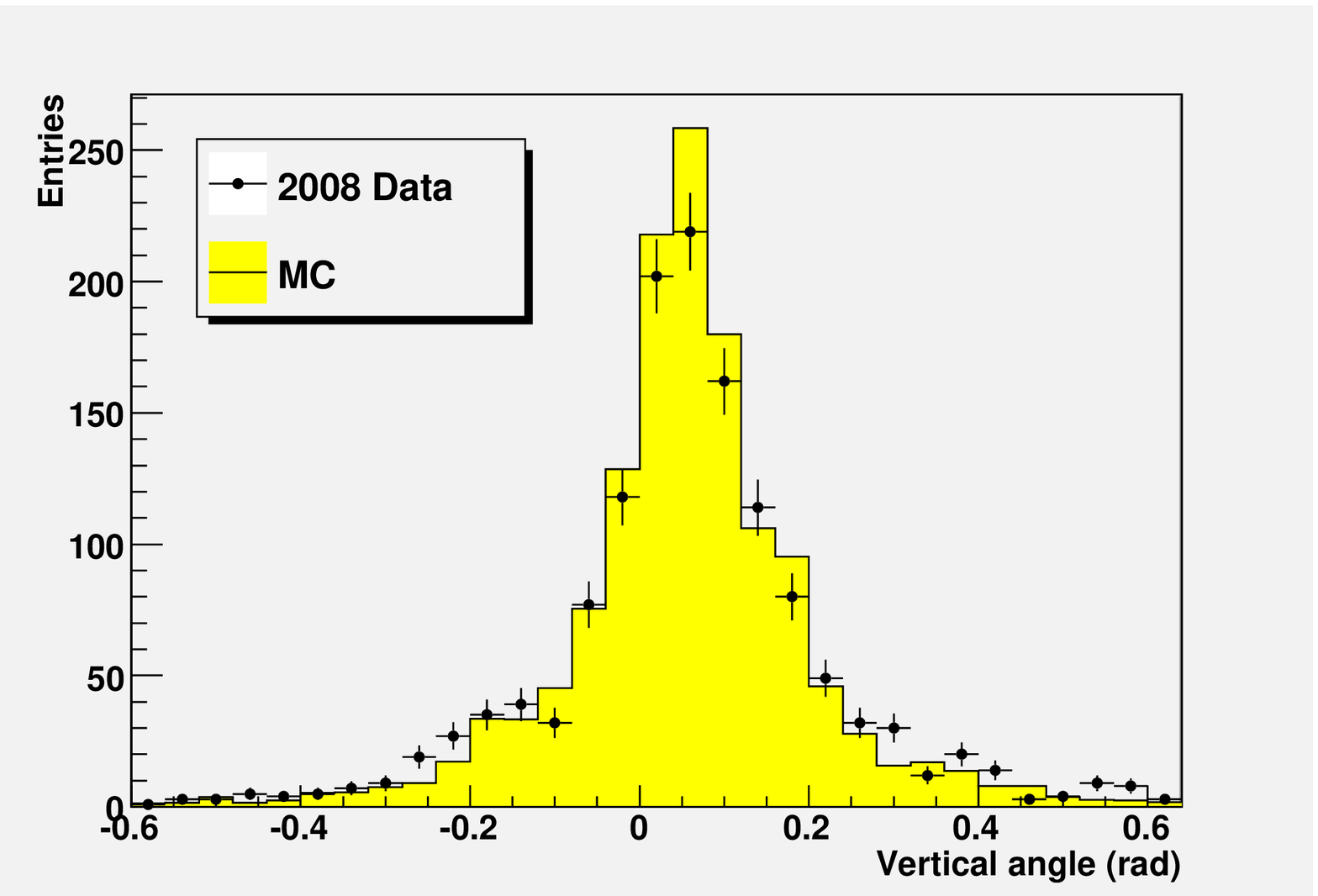}
  \caption{Left: momentum distribution of muons produced in CC neutrino interactions inside the OPERA target. Right: angular distribution of the muon tracks with respect to the horizontal axis.}
\label{fig:pmu_contained}
\end{center}
\end{figure}

The muon momentum distribution for events classified as CC interactions in the target is shown in the left panel of Figure \ref{fig:pmu_contained}. The distribution of the muon angle with respect to the horizontal axis is shown in the right panel of Figure \ref{fig:pmu_contained}; the beam direction angle is found to be tilted by 58 mrad, as expected from geodesy.

%
%


An extensive study of the beam monitoring is being performed by using neutrino interactions both in the whole OPERA detector and in the surrounding rock material. This will be the subject of a forthcoming publication.

\section{Combined analysis of electronic detectors and nuclear emulsion film data}

We describe in the following the breakdown of the different steps carried out to analyze neutrino interaction events from the identification of the ''fired'' brick up to the detailed kinematical analysis of the vertex in the emulsion films.

Once a trigger in the electronic detectors is selected to be compatible with an interaction inside a brick
the following procedure is applied \cite{detpap}:

\begin{enumerate}
  \item electronic detector data are processed by a software reconstruction program that selects the brick with the highest probability to contain the neutrino interaction vertex;
  \item this brick is removed from the target wall by the BMS and exposed to X-rays for film-to-film alignment. There are two independent X-ray exposures: the first one ensures a common reference system to the CS film doublet and the most downstream film of the brick (frontal exposure); the second one produces thick lateral marks on the brick edges, used for internal alignment and film numbering within the brick;
  \item after the first X-ray exposure the CS doublet is detached from the brick and developed underground, while the brick is kept in a box made of 5 cm thick iron shielding to reduce the radioactivity background;
  \item if the CS scanning detects tracks compatible with those reconstructed in the electronic detectors the second X-ray exposure (lateral marking) is performed and  the brick is brought to the surface laboratory. The brick is then exposed to cosmic-rays for about 24 hours in a dedicated pit in order to select high-energy cosmic muons to provide straight tracks for a refined (sub-micrometric) film-to-film alignment;
  \item the brick emulsion films are then developed and dispatched to the various scanning laboratories in Europe and Japan.
\end{enumerate}

\begin{figure}
\begin{center}
   \includegraphics[width=10cm]{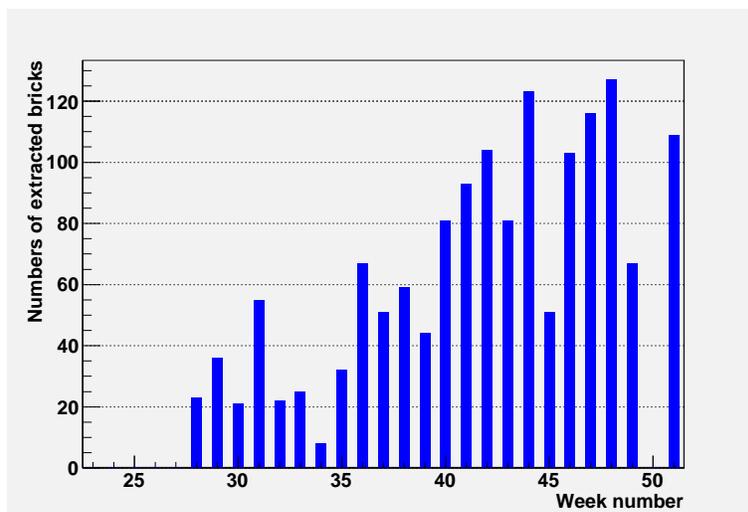}
  \caption{The number of bricks extracted per week by the BMS in 2008.} \label{bms}
\end{center}
\end{figure}

The procedure described above has proven to be successful. As an example, in Figure \ref{bms} we show the number of bricks extracted by the BMS per week, about one hundred. This is matched by the 100 CS developed and scanned per week in the LNGS (Italy) and Tono (Japan) scanning stations.

\vskip 30pt
\underline{\em Brick Finding and Changeable Sheet Interplay}
\vskip 10pt

 The efficiency for selecting the ''fired'' brick is the convolution of several effects and measurements. Here we discuss the two most important ones, the Brick Finding procedure and the Changeable Sheet measurement, for which preliminary results have been obtained from the analysis of partial samples of already scanned events.

The brick finding algorithm exploits the tracking capabilities of the OPERA electronic detectors and, by combining this information with the output of a Neural Network for the selection of the most probable wall where the interaction occurred, provides a list of bricks with the associated probability  that the interaction occurred therein. A preliminary estimate of the brick finding efficiency, limited to the extraction of the first most probable brick (for about 700 events) and not considering the small fraction ($<5\%$) of events for which the present electronic detector reconstruction fails, is compatible with the Monte Carlo estimate of $70\%$ computed for a standard mixture of CC and NC events. A higher efficiency can be obtained by extracting also bricks ranked with lower probabilities.

The  tracking efficiency  of single emulsion films can be measured by an exposure to high-energy pion beams and amounts to about 90\% \cite{ees2}. However, the measurement of the CS doublet efficiency in situ, in the OPERA detector, is by far more challenging, given the coarse resolution in the extrapolation of tracks from the electronic detectors to the CS.

At present, we are studying the CS tracking efficiency by two independent approaches:
 (a) all tracks produced in already located neutrino vertices are followed downstream and searched for in the corresponding CS doublet; (b) muon tracks reconstructed by the electronic detectors and found in the CS are properly normalized to the total number of CC events where at least one track (not necessarily the muon) is found in the CS. The two methods yield a preliminary efficiency for finding a track in both films of the CS doublet which is compatible with the conservative expectation of $90\%$ on a single film \cite{ees2}. The experimental efficiency has been evaluated on a sample of 100 events scanned in both the European and the Japanese laboratories. We are presently working in order to further increase this efficiency by employing more advanced analysis techniques.



\vskip 10pt
\underline{\em Vertex analysis}
\vskip 10pt

\begin{figure}
\begin{center}
  \includegraphics[width=10cm]{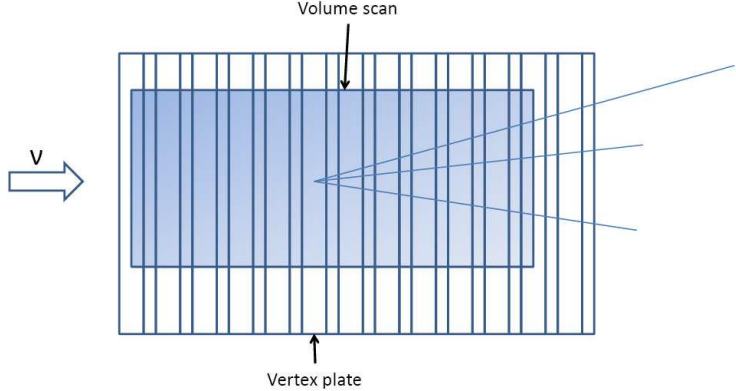}
  \caption{Schematic view of the volume scan performed around the stopping point of the track.}\label{figura1}
\end{center}
\end{figure}

All tracks measured in the CS are sought in the most downstream films of the brick and followed back until they are not found in three consecutive films. The stopping point is considered as the signature either for a primary or a secondary vertex. The vertex is then confirmed by scanning a volume with a transverse size of 1 cm$^2$ for 11 films in total, upstream and downstream of the stopping point (see Figure \ref{figura1}). Preliminary estimates of the vertex location efficiency are in agreement with the Monte Carlo expectations of $90\%$ and $80\%$ for CC and NC events, respectively. This evaluation has been performed on a sample of 500 located events.

\begin{figure}
\begin{center}
  \includegraphics[width=8cm]{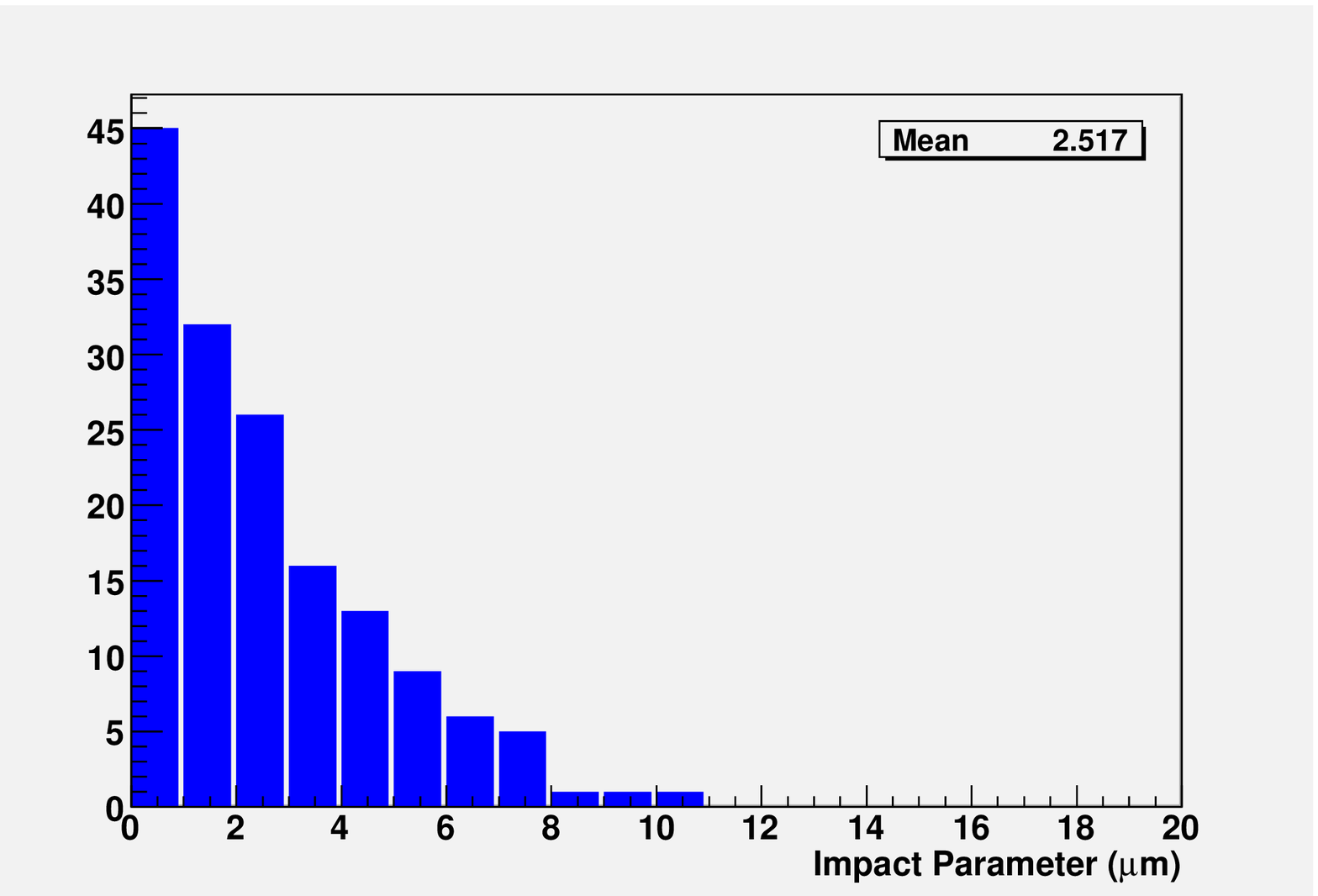}
  \includegraphics[width=8cm]{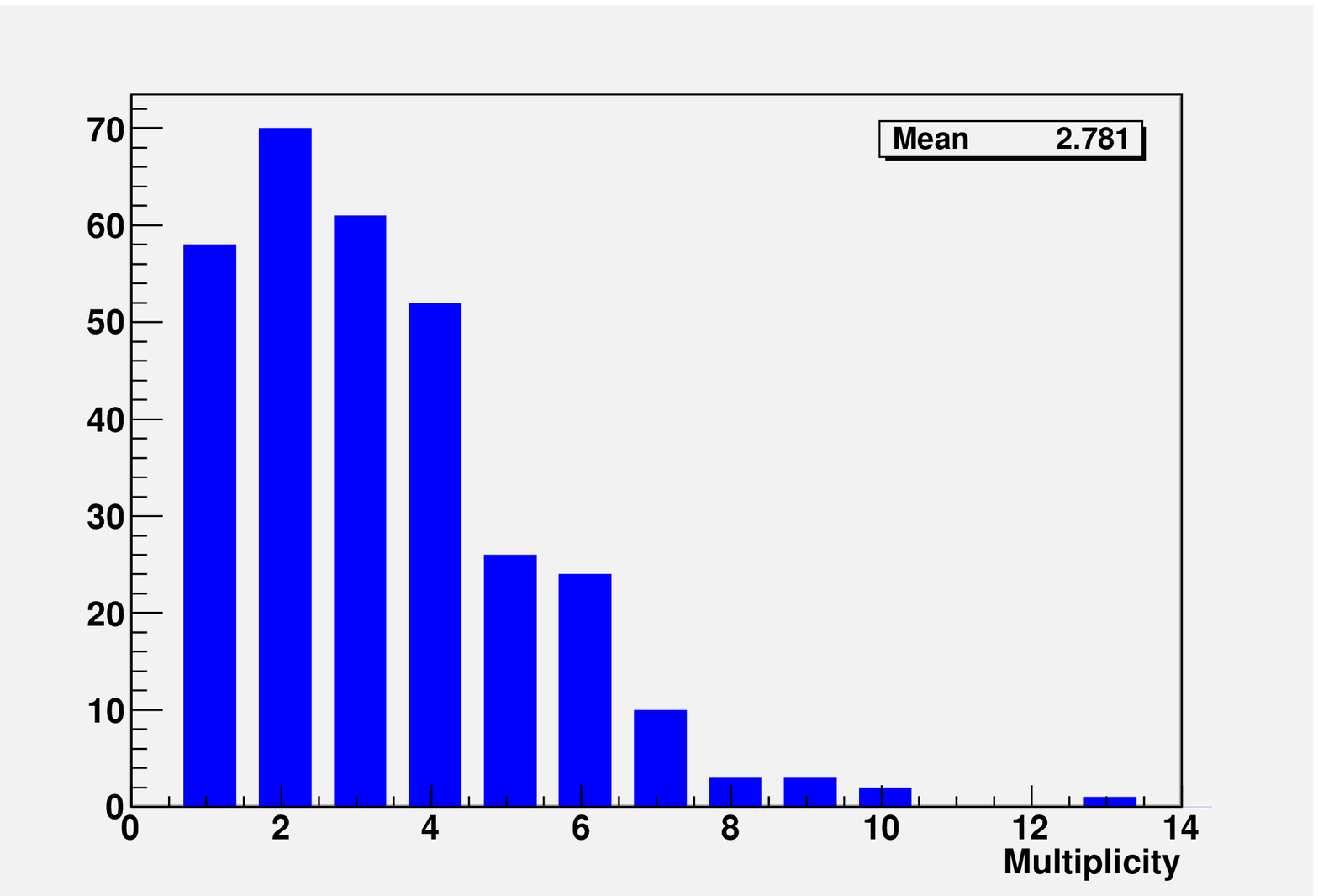}
  \caption{Left panel: impact parameter distribution of the muon track in CC events with respect to the reconstructed vertices. Right panel: charged track multiplicity distribution of the events.}\label{eve_rec1}
\end{center}
\end{figure}

The track impact parameter distribution of the muon in CC events with respect to the reconstructed vertex position and the event track multiplicity distribution are shown in Figure \ref{eve_rec1}. As expected, the impact parameter distribution is peaked at zero and has a mean value of 2.5 $\mu$m. The multiplicity distribution shows the anticipated enhancements for even track numbers due to the preferred interaction of neutrinos with neutrons.

\begin{figure}
\begin{center}
    \includegraphics[width=15cm]{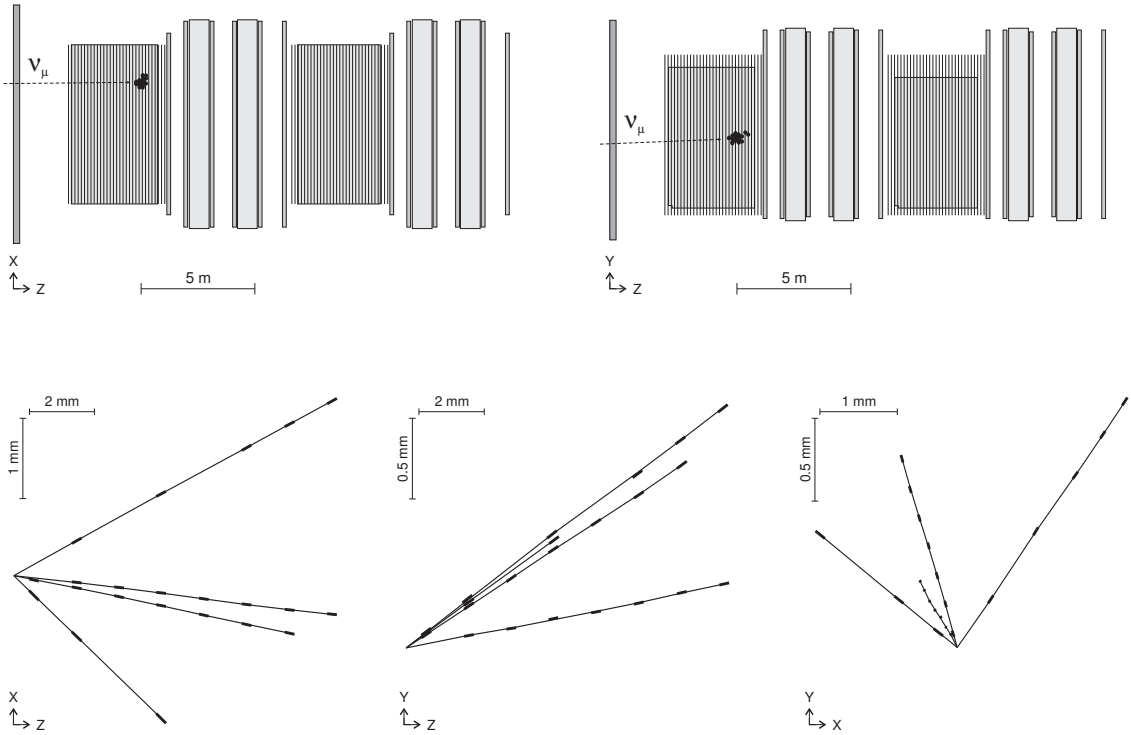}
  \caption{Top panels: online display of one NC event seen by the OPERA electronic detectors. The regions filled with bricks are highlighted. Bottom panels: the emulsion reconstruction is shown in the bottom panels: top view (bottom left), side view (bottom center), front view (bottom right).}\label{eve_NC}
\end{center}
\end{figure}

\begin{figure}
\begin{center}
    \includegraphics[width=15cm]{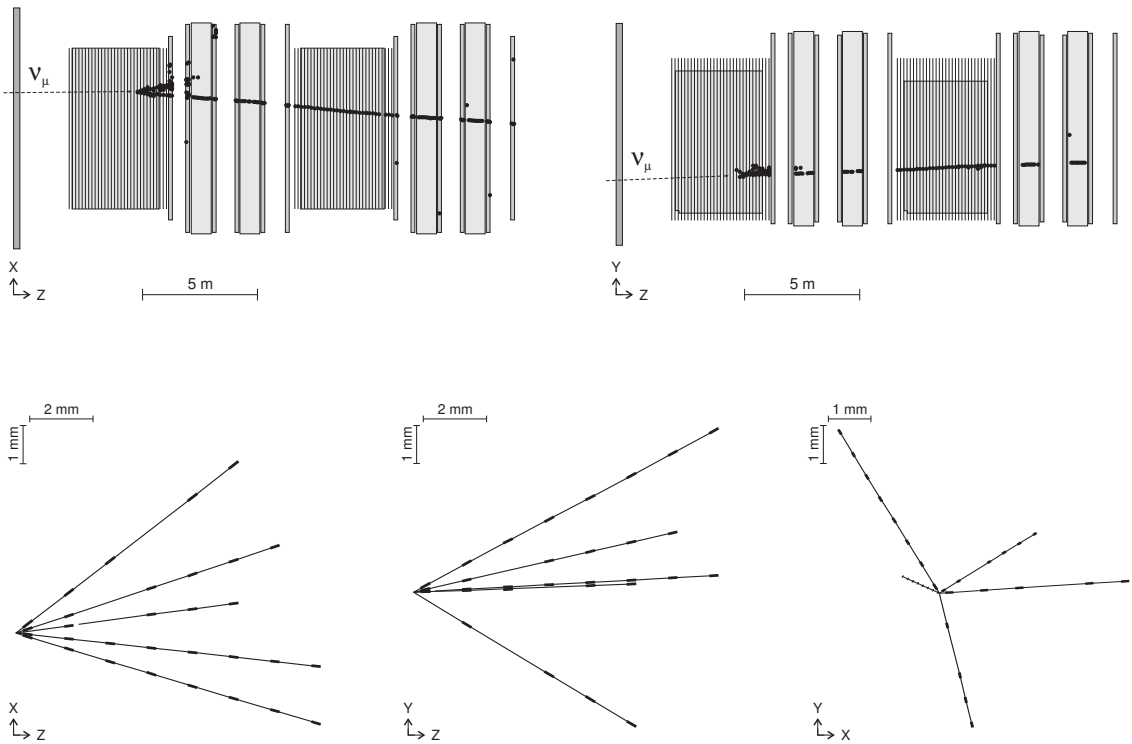}
  \caption{Top panels: online display of one CC event seen by the OPERA electronic detectors. Bottom panels: the emulsion reconstruction is shown in the bottom panels: top view (bottom left), side view (bottom center), front view (bottom right).}\label{eve_CC}
\end{center}
\end{figure}

\begin{figure}
\begin{center}
  \includegraphics[width=15cm]{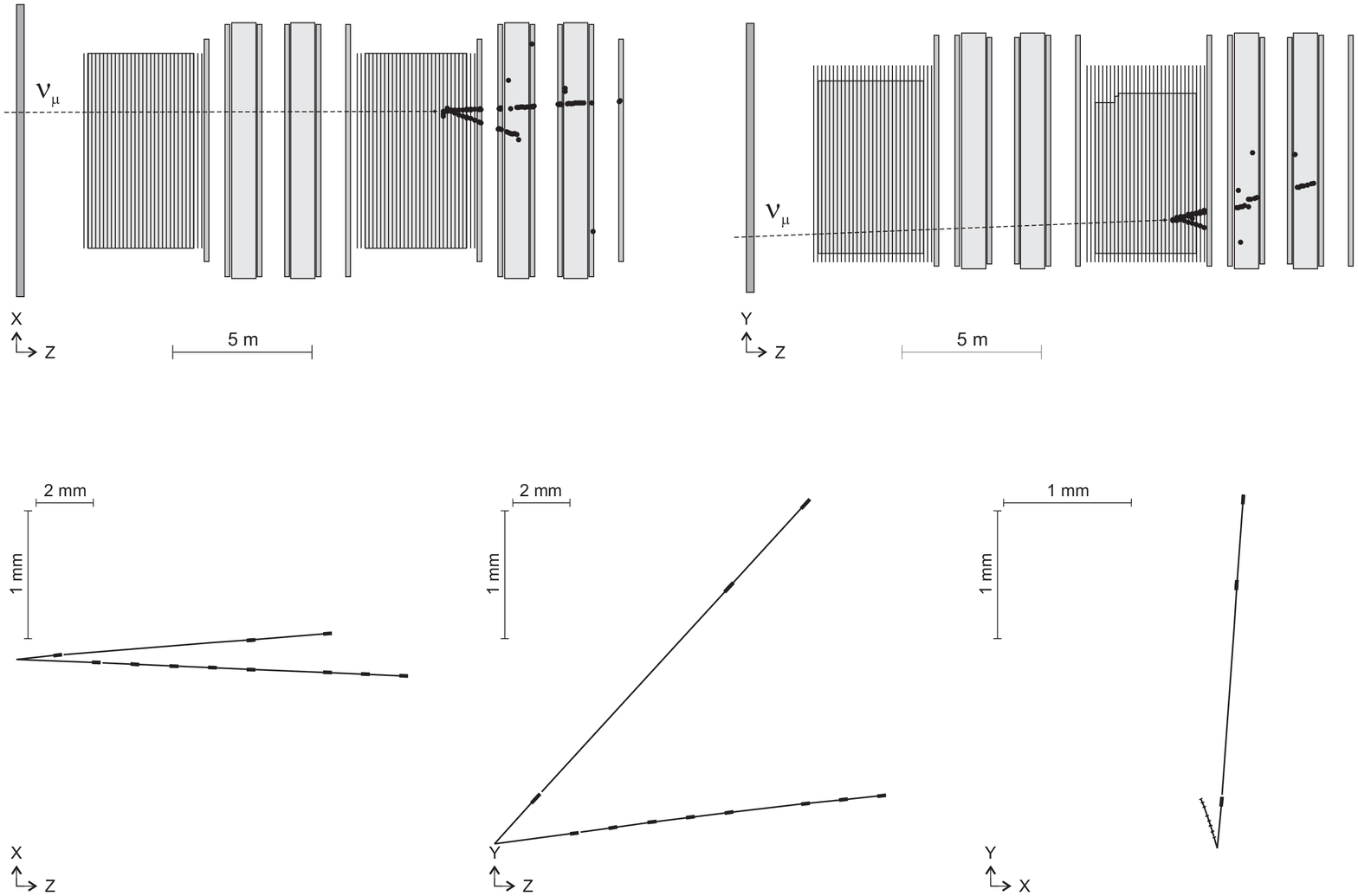}
  \includegraphics[width=7cm]{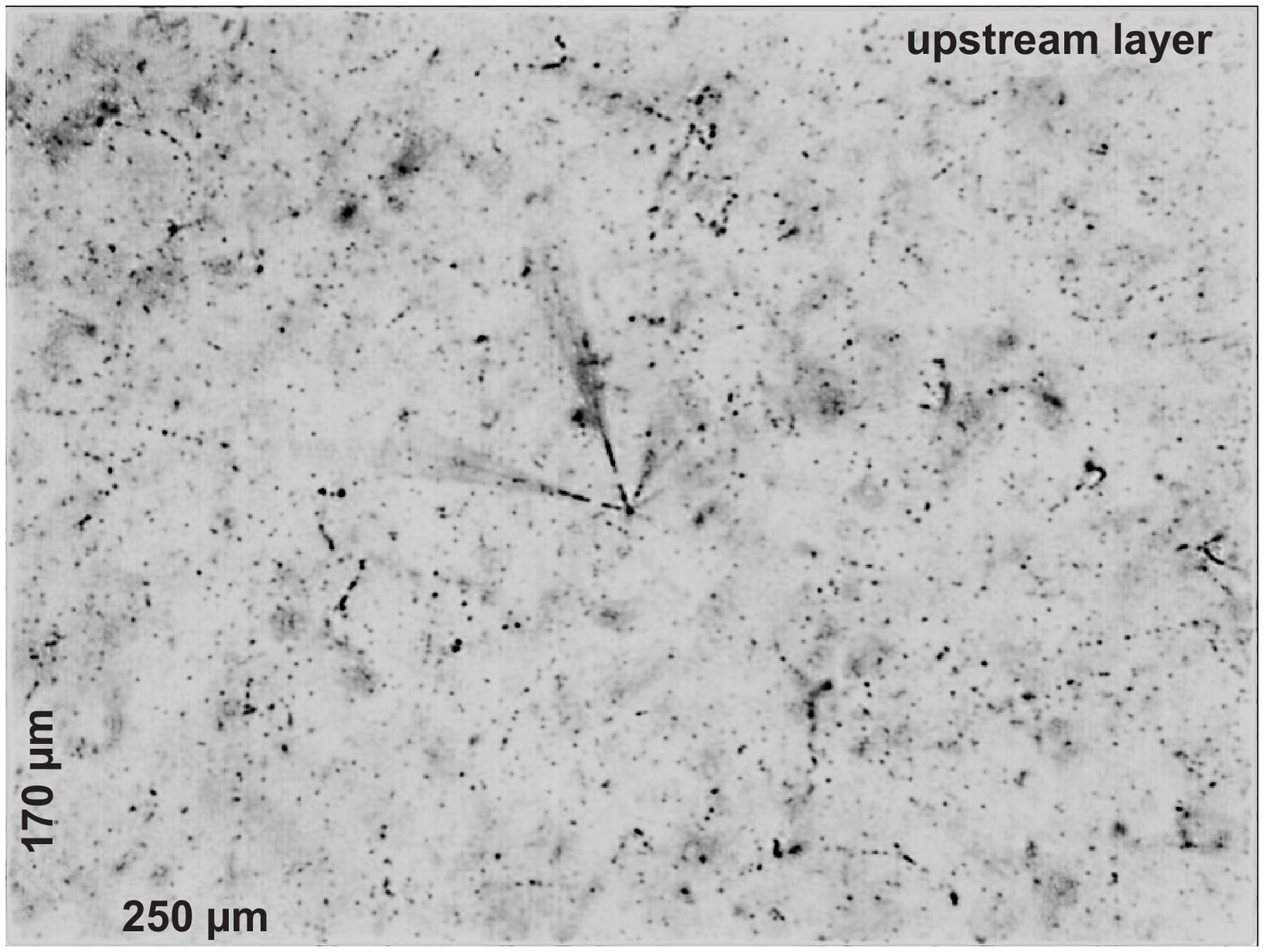}
   \hspace{1cm}
  \includegraphics[width=7cm]{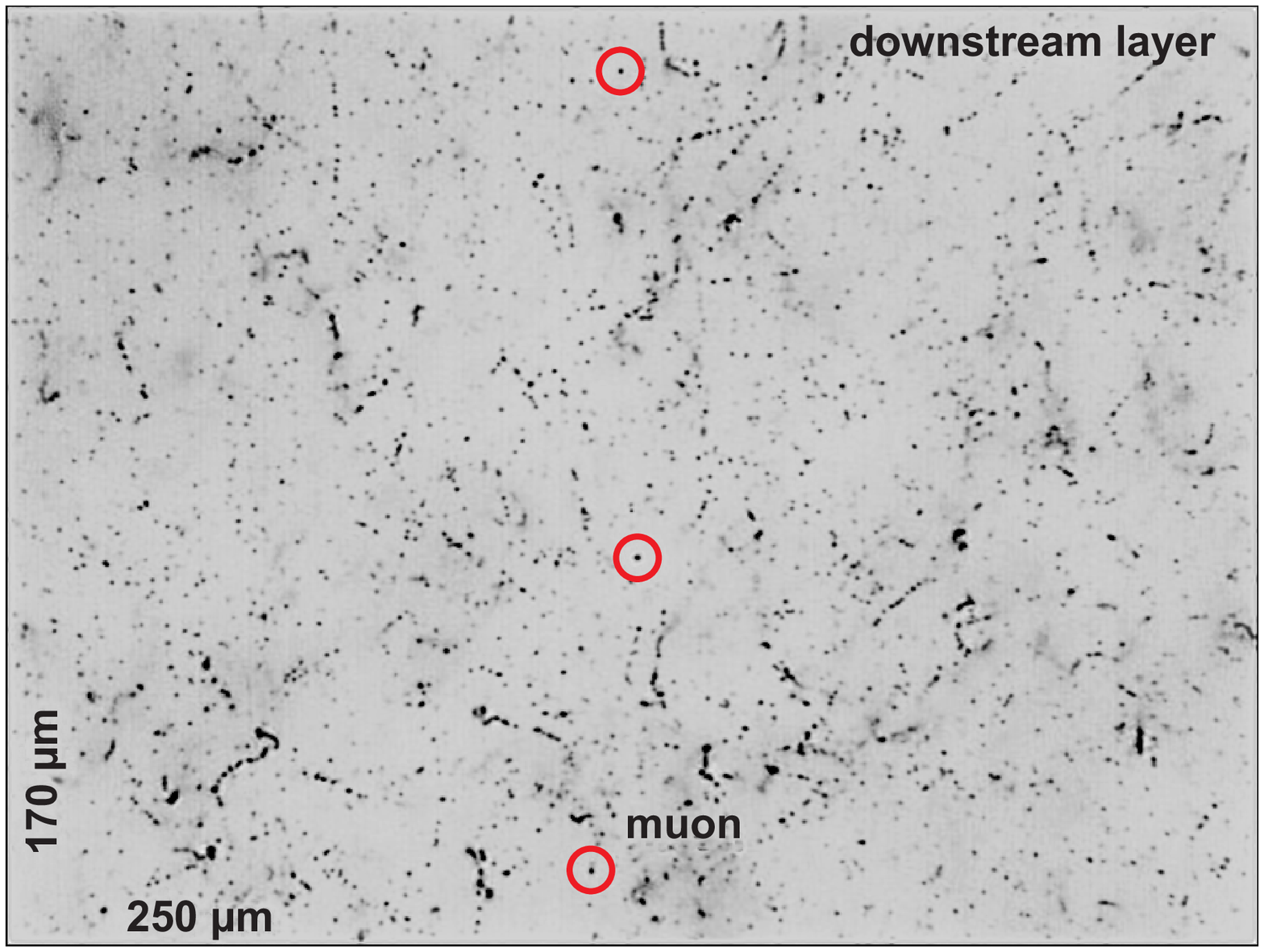}
  \caption{Display of the OPERA electronic detector of a $\nu_\mu$ CC interaction, top and side views. The emulsion reconstruction is shown in the middle panels: top view (bottom left), side view (bottom center), frontal view (bottom right). Bottom left panel : picture of the interaction vertex as seen by the microscope CMOS camera. The nuclear fragments produced in the interaction are visible. Bottom right panel : picture taken about 200 micron far from the interaction vertex. The minimum ionizing particles produced in the interaction are indicated by a circle. The muon track is also indicated.}\label{eve_pla}
\end{center}
\end{figure}

As an example, in Figures \ref{eve_NC} and \ref{eve_CC} we show a NC and a CC event, respectively, fully reconstructed in the brick. A very "peculiar" event is shown in Figure  \ref{eve_pla}: the neutrino  interaction occurred  in  the bottom  layer of an emulsion  film. Therefore,  the associated nuclear  fragments (large angle heavy ionizing tracks) are also visible in the film containing the vertex.

\vskip 10pt
\underline{\em Decay topologies}
\vskip 10pt


\begin{figure}
\begin{center}
  \includegraphics[width=10cm]{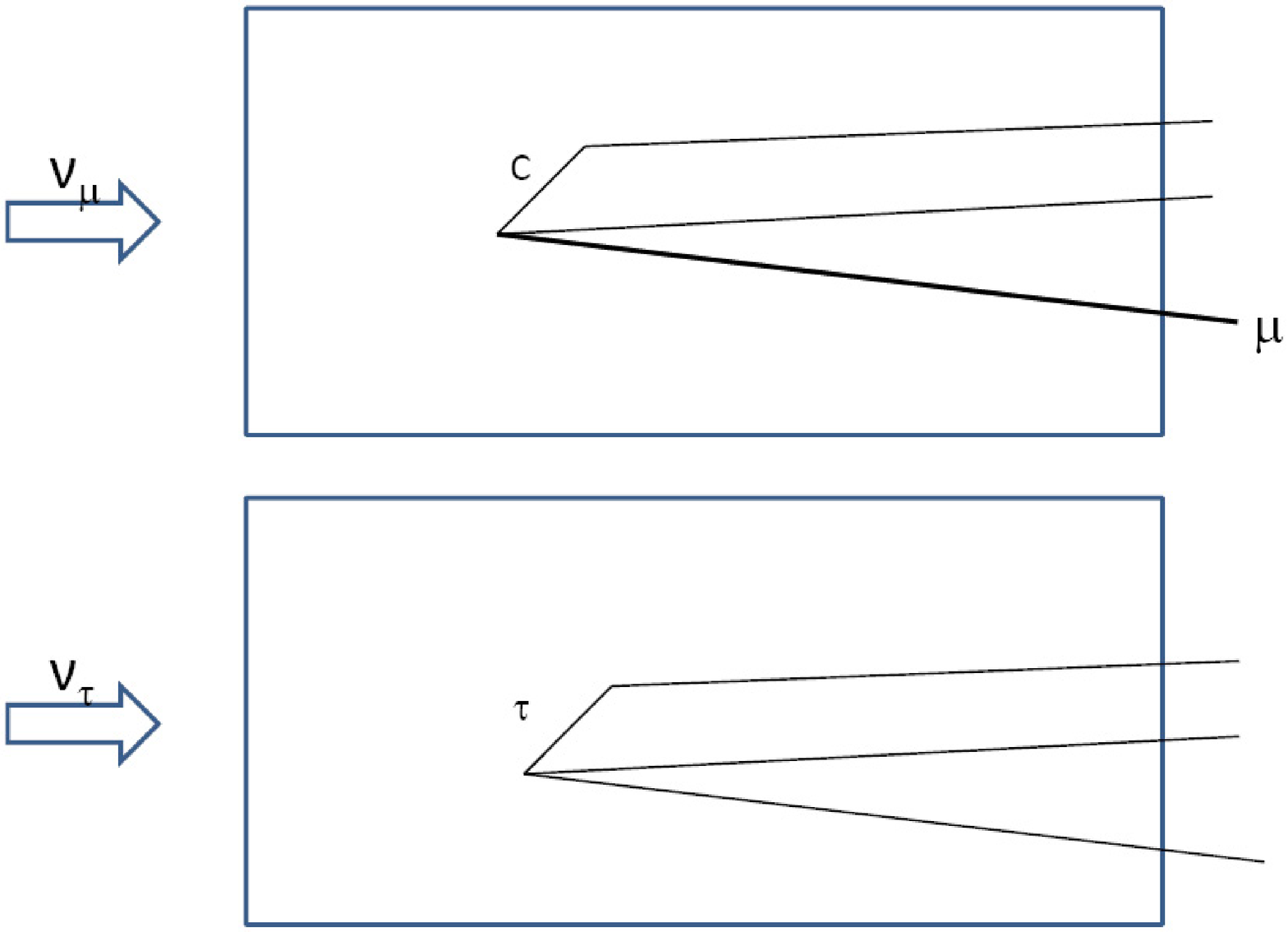}
  \caption{Schematic view of the charm and tau decay topologies.}\label{figura2}
\end{center}
\end{figure}

Charm production and decay topology events have a great importance in OPERA for two main reasons. On the one hand in order to certify the observation of $\tau$ events one should prove the ability of observing charm events at the correct expected rate. On the other hand, since charm decays exhibit the same topology as $\tau$ decays, they are a potential source of background if the muon at the primary vertex is not identified (see Figure \ref{figura2}). Therefore, searching for charm-decays in events with the primary muon correctly identified provides a direct measurement of this background.

Charm decay topologies were searched for in the sample of located neutrino interactions. Two events with charm-like topologies were found. By using the neutrino-induced charm-production cross-section measured by the CHORUS experiment \cite{chorus} about 3 charged-charm decays are expected to be observed in the analyzed sample.

\begin{figure}
\begin{center}
  \includegraphics[width=15cm]{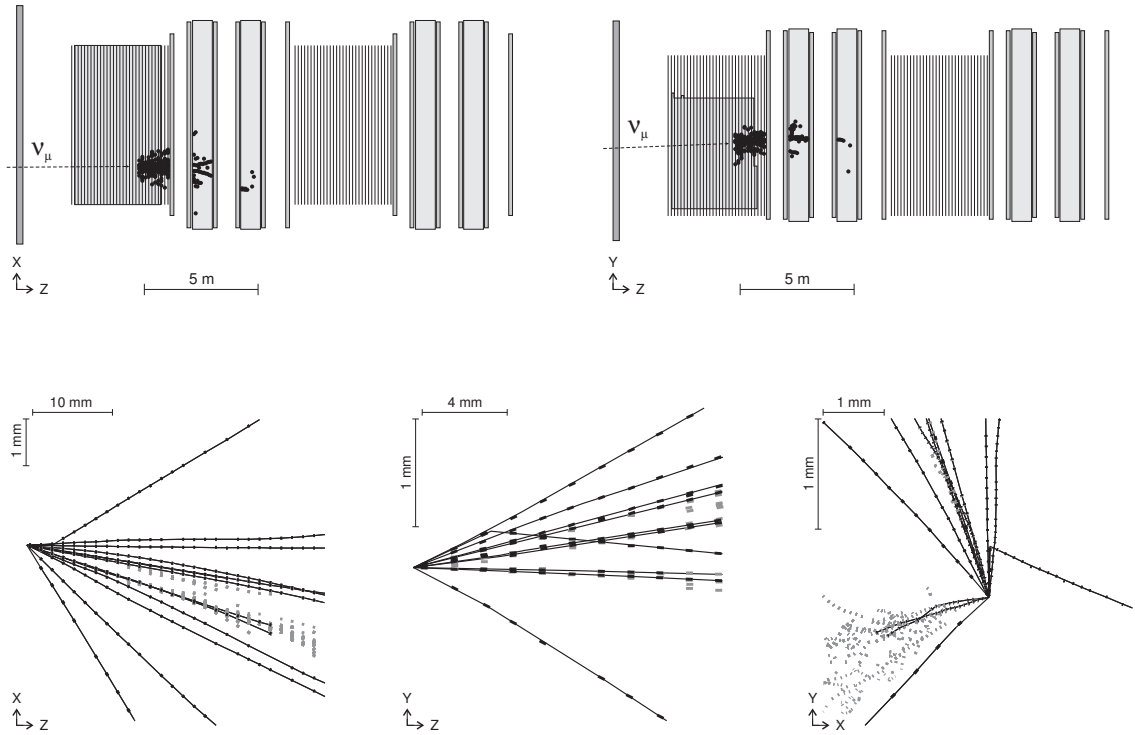}
  \vspace{1cm}
  \caption{Online display of the OPERA electronic detector of a $\nu_\mu$ charged-current interaction with a charm-like topology (top panel). The emulsion reconstruction is shown in the bottom panels where the charm-like topology is seen as a track with a kink: top view (bottom left), side view (bottom center), frontal view (bottom right).}\label{eve_na}
  \end{center}
\end{figure}

The event in Figure \ref{eve_na} has high track multiplicity at the primary vertex and one of the scan-back tracks
shows a kink topology. The measured decay angle is 204 mrad and the flight length of the decaying particle is 3247 $\mu$m. The decay occurred in the third lead plate downstream of the interaction plate. No large angle tracks are produced at the decay vertex. This allows to further rule out the hadronic interaction hypothesis. The muon track and the charm candidate track lie in a back-to-back configuration ($\Delta \phi$ $\simeq$ 165$^\circ$) as one would expect for charm production.
The daughter momentum, measured by using the Multiple Coulomb Scattering technique is
$3.9^{+1.7}_{-0.9}$ GeV/c at the $90\%$ C.L. Therefore, at the $90\%$ C.L. the transverse momentum ranges between 600 MeV/c and 1150 MeV/c, well above the cut of 250 MeV/c applied to reject hadronic decays. According to the FLUKA Monte Carlo \cite{fluka} the probability that a hadron interaction mimics a charm-decay with transverse momentum larger than 600 MeV/c is only $4\times10^{-4}$.

\begin{figure}
\begin{center}
  \includegraphics[width=15cm]{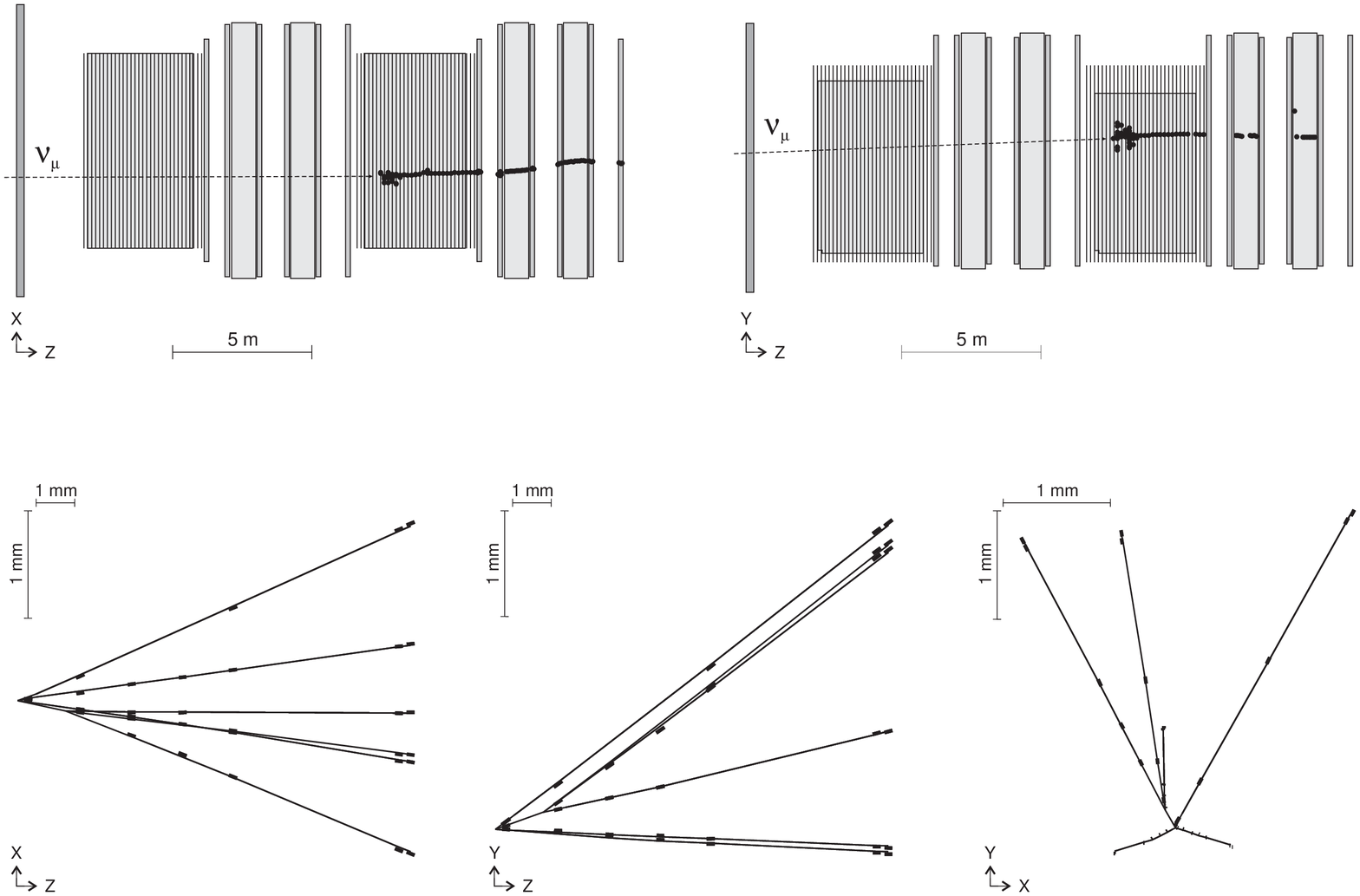}
  \vspace{1cm}
  \caption{Online display of the OPERA electronic detector of a $\nu_\mu$ charged-current interaction with a charm-like topology (top panel). The emulsion reconstruction is shown in the bottom panels where the charm-like topology is seen as a three-prongs secondary vertex: top view (bottom left), side view (bottom center), frontal view (bottom right).}\label{eve_pd}
  \end{center}
\end{figure}

The second charm-like topology is shown in Figure \ref{eve_pd}. A 4-prong primary vertex is observed originating at a depth
of about 30 $\mu$m in the upstream lead plate. The charmed hadron track points to a 3-prong
decay vertex located at a distance of 1150 $\mu$m from the primary vertex (200 $\mu$m inside the lead). All tracks have a clear CS tag.
The interaction occurs downstream in the brick and the tracks only cross  four emulsion films and the CS doublet (the two most downstream hits in the figure). The muon track and the charm candidate track lie in a back-to-back configuration ($\Delta \phi$ $\simeq$ 150$^\circ$). The relativistic $\gamma$ of the charmed parent has been roughly estimated as the inverse of the average angle in space that the daughter tracks form with it. This leads to a $\gamma$ value of about 8.6, implying a parent high momentum of $\sim16$ GeV/c (assuming the $D^+$ mass). The momenta of the daughter tracks have also been measured by extracting the downstream brick and using the Multiple Coulomb Scattering technique. The measured values are $p_1$=$2.4^{+1.3}_{-0.6}$, $p_2$=$1.3^{+0.4}_{-0.3}$ and $p_3$=
$1.2^{+1.7}_{-0.4}$ GeV/c (transverse momenta of about 610, 90 and 340 MeV/c, total momentum: $4.8^{+2.2}_{-0.8}$ GeV/c), at the $90\%$ C.L. The probability of a
hadron interaction has been evaluated using FLUKA and amounts to $10^{-6}$. Assuming a $D \to K\pi\pi$ decay, an invariant mass of 1.1$^{+0.2}_{-0.1}$ GeV/c$^2$ is obtained. On the other hand assuming a $D_s \to K K \pi$ decay an invariant mass of 1.5$^{+0.4}_{-0.1}$ GeV/c$^2$ is derived. In the latter case the invariant mass is consistent with the mass of a charmed hadron while in the second case the consistency is marginal. The probability of a decay in flight of a $K$ is about $10^{-3}$.


\section{Conclusions}

The 2007 and 2008 CNGS runs constitute an important milestone for the LNGS OPERA experiment searching for $\nu_\mu\rightarrow\nu_\tau$ oscillations. First samples of neutrino interaction events have been collected in the emulsion/lead target and allowed to check the complete analysis chain starting from the trigger down to the neutrino vertex location in the emulsions and to the topological and kinematical characterization of the event.

In this paper we reported on the capability in performing an online identification, extraction and development of the bricks where the neutrino interaction occurred, vertex location and kinematical reconstruction. The overall performance of the experiment during the running phase and through the analysis chain can be summarized by stating that:

\begin{itemize}
  \item all electronic detectors performed excellently allowing the precise localization of the brick hit by the neutrino;
  \item the electronic detector event reconstruction was tuned to the brick finding procedure which operated  for the first time with real neutrino events providing good results;
  \item all experimental activities from brick removal upon identification to the X and cosmic-ray exposures, brick disassembly and emulsion development, have been
  successfully accomplished. At present more than 100 bricks per week can be routinely handled;
  \item the scanning of the Changeable Sheets can be performed with the expected detection efficiencies;
  \item vertex location was successfully attempted for both CC and NC events;
  \item the topological and kinematical analyses of the vertices were successfully exploited and led to an unambiguous interpretation of neutrino interactions. In particular two events with a charm-like topology were found so far in the analyzed sample. This is fully consistent with expectations based on the known neutrino-induced charm production cross-section.
\end{itemize}

The above considerations make us confident that the OPERA experiment is definitely in its production phase with the CNGS beam and that the scene has been set for the discovery of the $\tau$ appearance.

\indent


\section {Acknowledgements}

We thank CERN for the commissioning of the CNGS facility and
for its successful operation, and INFN for the continuous support given
to the experiment during the construction, installation and commissioning phases through its LNGS laboratory.
We warmly acknowledge funding from our national agencies: Fonds de la Recherche Scientifique - FNRS and Institut Interuniversitaire des Sciences Nucleaires for Belgium, MoSES for Croatia, IN2P3-CNRS for France, BMBF for Germany, INFN for Italy, the Japan Society for the Promotion of
Science (JSPS), the Ministry of Education, Culture, Sports, Science and Technology (MEXT)
and the Promotion and Mutual Aid Corporation for Private Schools of Japan for Japan, SNF and
ETH Zurich for Switzerland, the Russian Foundation for Basic Research (grants 08-02-91005 and 08-02-01086) for Russia, the Korea Research Foundation Grant (KRF-2007-013-C00015) for Korea. We are also indebted to INFN for providing fellowships and
grants to non Italian researchers. Finally, we are indebted to our technical collaborators for the excellent quality of their work over many years of design, prototyping and construction of the detector and of its facilities.

\end{document}